\documentclass[9pt, sigconf, authorversion]{acmart}

\usepackage{microtype}
\usepackage{amsmath,amsfonts}
\usepackage{graphicx}
\usepackage{textcomp}
\usepackage{xcolor}
\usepackage{comment}
\usepackage{tikz}
\usetikzlibrary{decorations.pathmorphing}

\usepackage{pifont}
\usepackage{booktabs}
\usepackage{multirow}

\PassOptionsToPackage{hyperfootnotes=true}{hyperref}

\usepackage[linesnumbered,ruled,vlined]{algorithm2e}
\usepackage{bbold}

\usepackage{tabularx}
\usepackage{adjustbox}
\usepackage{tcolorbox} 
\usepackage{newunicodechar}
\newunicodechar{⁡}{}

\usepackage{array, makecell} 
\usepackage{multirow}
\usepackage{makecell}

\usepackage{hhline}
\usepackage{booktabs}

\SetCommentSty{mycommfont}
\usepackage{lipsum} 
\usepackage{textcomp}

\usepackage{tikz}
\usetikzlibrary{fit,calc}

\colorlet{pink}{red!40}
\colorlet{blue}{cyan!60}

\ExplSyntaxOn
\NewDocumentCommand \tempfnlabel { m }
{
    \int_zero_new:c { g_tmpfncounter_#1_int }
    \int_gset:cn { g_tmpfncounter_#1_int } { \value{footnote } }
}

\NewDocumentCommand \tempfncounter { m }
{
    \int_use:c { g_tmpfncounter_#1_int }
}
\ExplSyntaxOff

\newcommand{\ignore}[1]{}
\AtBeginDocument{%
  \providecommand\BibTeX{{%
    \normalfont B\kern-0.5em{\scshape i\kern-0.25em b}\kern-0.8em\TeX}}}
\setcopyright{acmlicensed}
\acmConference[DAC '24]{DAC '24: ACM Design Automation Conference}{June 23--27,
  2024}{San Francisco, CA, USA}
\acmDOI{10.1145/3649329.3655953}

\begin{document}

\title{Token-Picker: Accelerating Attention in Text Generation with Minimized Memory Transfer via Probability Estimation}



\author{Junyoung Park$^{1,2}$, Myeonggu Kang$^1$, Yunki Han$^1$, Yanggon Kim$^{2}$, Jaekang Shin$^{1,2}$, Lee-Sup Kim$^1$} 

\affiliation{
 \institution{$^1$Korea Advanced Institute of Science and Technology, $^2$System LSI, Samsung Electronics}
 \institution{\textit{\{wnsdud9731, jaekangshin\}@gmail.com, \{mgkang95, yunki.han, leesup\}@kaist.ac.kr, yanggon.kim@samsung.com}}
 \country{South Korea}
}

\renewcommand{\shortauthors}{Junyoung Park et al.}



\begin{abstract}
The attention mechanism in text generation is memory-bounded due to its sequential characteristics. Therefore, off-chip memory accesses should be minimized for faster execution. Although previous methods addressed this by pruning unimportant tokens, they fall short in selectively removing tokens with near-zero attention probabilities in each instance. Our method estimates the probability before the softmax function, effectively removing low probability tokens and achieving an 12.1x pruning ratio without fine-tuning. Additionally, we present a hardware design supporting seamless on-demand off-chip access. Our approach shows 2.6x reduced memory accesses, leading to an average 2.3x speedup and a 2.4x energy efficiency.
\end{abstract}

\maketitle
\renewcommand{\thefootnote}{}
\footnotetext[1]{This work was done when J. Park and J. Shin were at KAIST.}

\section{Introduction} \label{sec:intro}
Text generation using Large Language Models \cite{gpt3, opt, llama2} have been instrumental in advancing applications such as chatbot systems and virtual assistants. With their growing importance, several companies are striving to integrate these applications into their hosted services, highlighting the importance of efficient inference.

Language models are based on the autoregressive transformer model specialized for text generation. At its core, there is an attention mechanism and a fully connected (FC) layer. The attention mechanism captures context within text sequences, while the FC layer uses weights to transform activations into higher representations. This setup sequentially creates words using prior ones to form complete sentences. However, this sequential property in text generation makes the workload memory-bound, not fully utilizing the computing resources of GPUs. To address this, recent research \cite{vllm, efficiently_scaling, orca} has enhanced throughput of generation inference by adopting dynamic batching, allowing multiple requests to share the weights. It enables to amortize the transferring cost and improves parallel processing. Still, the attention mechanism faces with memory challenges because each user's sequence remains separate. As more users are batched together, memory accesses demands rise. Thus, minimizing the memory transfer of attention can improve throughput and energy efficiency in batching scenario. 

Previous works \cite{spatten, sanger, leopard} could achieve this by leveraging the inherent redundancy by softmax, which turns token correlation scores into probabilities and often produces many near-zero values. 
However, these approaches overlooked the varying number of unimportant tokens across instances and could not selectively spare the important ones only.
Furthermore, retraining is often required for high pruning rate, as the distribution needs to be adjusted to fit each method.

To tackle the aforementioned problems, we introduce an adaptive token pruning method that aligns with each instance. Before completing all correlation calculations, our method estimates token probabilities and eliminates those below a set threshold. Memory accesses during the estimation is minimized by: 1) beginning with the initial bit chunk of a key vector, where a bit chunk refers to a segment of bits from the vector elements;  2) estimating the probability to decide on pruning or retrieving the next chunk;  and 3) if the chunk is needed, efficiently managing the access delay by executing independent computation. Notably, our method achieves a higher pruning ratio compared to previous techniques, without the need for retraining.

The \textbf{major contributions} of this paper are as follows:
\begin{itemize}
    \item We propose a probability estimation method to prune redundant tokens, showing 12.1$\times$ pruning ratio in average. 
    \item We present out-of-order score calculations to support on-demand DRAM request, further reducing 1.39$\times$ transfers.
    \item We design a tailored hardware to support the proposed method, showing 2.28$\times$ speedup and 2.41$\times$ energy efficiency.
\end{itemize}

\section{BACKGROUND \& MOTIVATION} \label{sec:background_motivation}
\subsection{Autoregressive transformer model} \label{subsec:background}

\subsubsection{Transformer architecture}

Language models utilize transfor-mer architecture. Its main components are the self-attention and the feed-forward network (FFN). Self-attention discerns correlations within token embeddings $(\textbf{x}_1,\ldots,\textbf{x}_n) \in \mathbb{R}^{n\times d}$ to capture context information. The specific operations within the self-attention are detailed in the following equations:

\vspace*{-\baselineskip}
\begin{gather}
\mathbf{q}_t,\hspace{0.2em}\mathbf{k}_t,\hspace{0.2em}\mathbf{v}_t = \mathbf{W}_q \mathbf{x}_t,\hspace{0.2em}\mathbf{W}_k \mathbf{x}_t,\hspace{0.2em}\mathbf{W}_v \mathbf{x}_t \label{eq_1}\\
s_{ti} = \mathbf{q}_t^\intercal \cdot \mathbf{k}_i/\sqrt{d_h}\hspace{0.2em},\hspace{0.8em}p_{ti} = \dfrac{exp(s_{ti})}{\sum_{k=1}^{t}exp(s_{tk})} \quad \text{for } 1 \le i \le t \label{eq_2}\\
\mathbf{o}_t = \sum_{i=1}^{t}p_{ti} \mathbf{v}_i \label{eq_3}
\end{gather}

(\ref{eq_1}) Initially, the embedding vector ($\mathbf{x}_t$) is multiplied by weights $(\mathbf{W}_q, \mathbf{W}_v, \mathbf{W}_k)$ to transform into query, key, and value vector (each $\mathbf{q}_t, \mathbf{k}_t, \mathbf{v}_t$). Following this, (\ref{eq_2}) the query and key vectors create correlation scores $(s_{ti})$ through scaled dot-product. Note that in autoregressive models, the positions for the keys precede the query. After that, the scores $(s_{t1},\ldots, s_{tt})$ are normalized into attention probabilities $(p_{t1},\ldots, p_{tt})$ using the softmax function, indicating strength of association between token $x_t$ and $x_i$. (\ref{eq_3}) Finally, using these probabilities as weights, value vectors are multiplied and summed to produce attention output $(\mathbf{o}_t)$. Subsequent operations include FFN layer, layer normalization, and residual connection to finalize one layer. Multiple iterations through subsequent layers followed by the final output embedding yield one token generation.

\begin{figure}[t]
  \centering
  \vspace*{-3mm}
  \includegraphics[width=\linewidth]{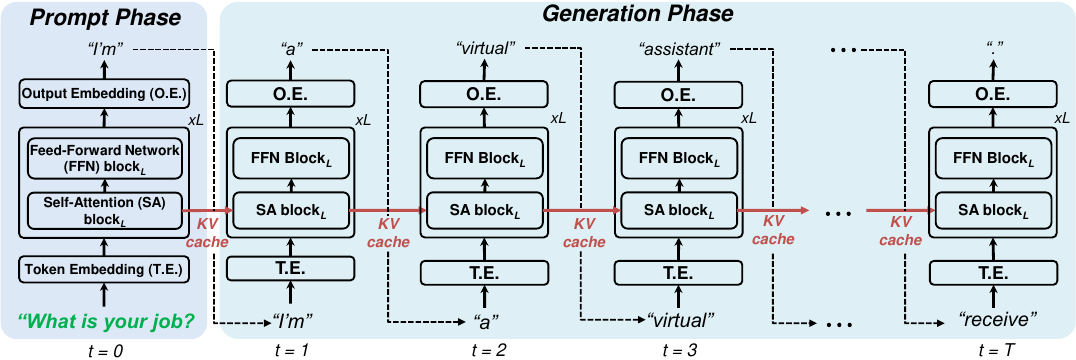}
  \vspace*{-7mm}
  \caption{Transformer-based autoregressive text generation.}
  \label{fig:text-generation}
  \Description{Generating words based on input prompts "What is your job?".}
  \vspace*{-\baselineskip}
\end{figure}

\subsubsection{KV caching} \label{subsubsec:KV_caching}
As illustrated in Fig. \ref{fig:text-generation}, the text generation process consists of the prompt phase and generation phase. During the prompt phase, an entire prompt sequence $(x_1,\ldots,x_n)$ is used to predict a new token $x_{n+1}$ (e.g., "I’m"). Within this phase, $n$ sets of query, key, and value are produced and utilized for the self-attention block of each layer. Meanwhile, the generated keys $(\mathbf{k}_1,\ldots,\mathbf{k}_n)$ and values $(\mathbf{v}_1,\ldots,\mathbf{v}_n)$ are stored to prevent redundant creation in the following token generation. This technique is known as \textit{KV caching}.

In the generation phase, tokens are sequentially generated until the maximum sequence length or an end-of-sequence token (<$eos$>) is encountered. At a given time $t$, the model takes an input token $x_{n+t}$ to produce the following token $x_{n+t+1}$. The input is constructed as a vector from a single token, leading to the execution of a General Matrix-Vector Multiplication (GEMV) operation that makes the workload memory-bound. Therefore, latency and energy consumption are dominated by off-chip access in this phase.

\subsection{Motivation} \label{subsec:motivation}

\subsubsection{Memory transfer overhead}

\begin{figure}[t]
  \centering
  \includegraphics[width=\linewidth]{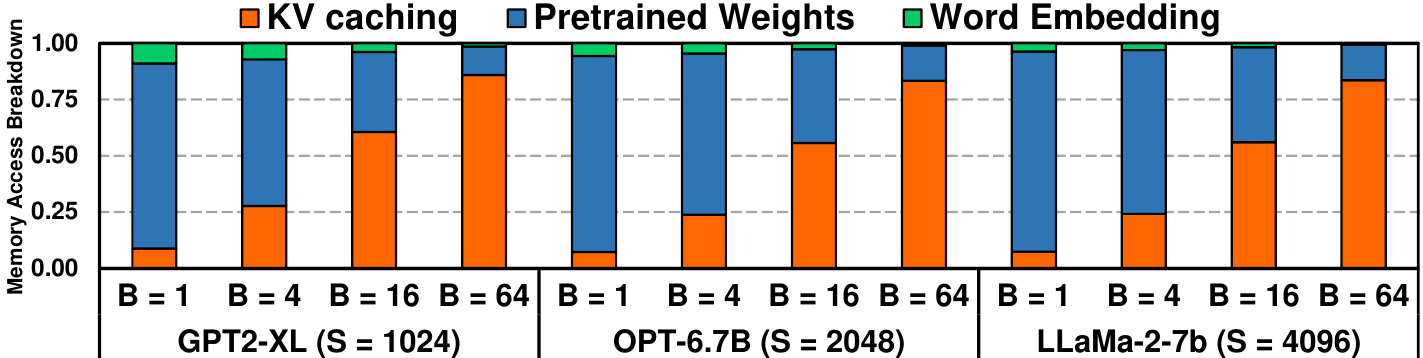}
  \vspace{-5mm}
  \caption{Memory transfer breakdown.}
  \Description{A woman and a girl in white dresses sit in an open car.}
  \label{fig:transfer_breakdown}
  \vspace{-5mm}
\end{figure}

Efficient batching technique was developed to improve the generation process. It spreads out the cost of loading pre-trained weights across multiple requests, thereby increasing inference efficiency. Yet, the unique $KV$ cache in each self-attention cannot be shared, resulting in increased memory transfer overheads. Fig. \ref{fig:transfer_breakdown} shows the breakdown of off-chip memory accesses during the generation phase for different batch sizes, all set to each model's maximum context length. While $KV caching$ transfer is 7.8\% at a batch size of 1, it becomes 84.3\% for a batch size of 64, leading to prolonged generation latency as demonstrated in \cite{efficiently_scaling}. Thus, minimizing latency in self-attention is essential to handle larger batch sizes for enhancing the benefits from weight sharing. This emphasizes the need to reduce off-chip accesses for \textit{KV} transfers in self-attention.

\begin{figure}[t]
  \centering
  \includegraphics[width=\linewidth]{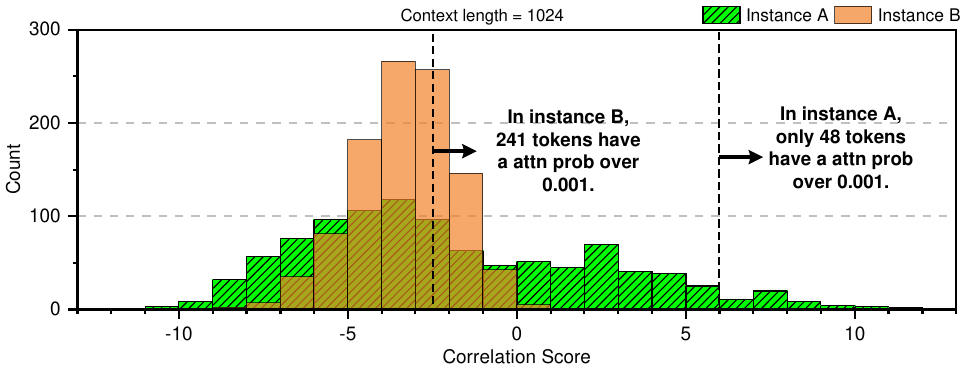}
  \vspace{-7mm}
  \caption{Various attention score distribution.}
  \Description{A woman and a girl in white dresses sit in an open car.}
  \label{fig:distribution}
  \vspace{-7mm}
\end{figure}

\subsubsection{Distribution-aligned pruning} \label{Sec}

The softmax operation in self-attention amplifies differences among correlation scores exponentially, resulting in numerous tokens with near-zero $p_i$ in Eq. (\ref{eq_3}). Thus, the corresponding value vectors $\mathbf{v}_i$ barely affect the attention output $\mathbf{o}_t$ and can be pruned with minimal performance loss. Using this property, there have been several works \cite{spatten, sanger, leopard} to lighten the cost of self-attention block. Among those, SpAtten \cite{spatten} reduced the access for \textit{KV cache} with cascade token/head pruning and local value pruning, which retains tokens with the highest probability at a pre-defined ratio. While the method is effective in reducing \textit{KV} transfers, it often overlooked variations in the number of unimportant tokens across instances. 

Fig. \ref{fig:distribution} demonstrates the variability by comparing the number of dominant tokens (i.e., probability over $10^{-3}$) in identical setups—same layer, head, and context length. In the two instances, only 4.6\% of tokens in instance A exceed a probability of $10^{-3}$, while 23.5\% in instance B. This discrepancy arises from the relative nature of the softmax; wider distribution of scores, with greater differences between scores, leads to fewer dominant tokens. Therefore, fixed-ratio pruning strategy could either remove dominant or keep unimportant tokens, undermining the overall performance and efficiency. Furthermore, it necessitates additional fine-tuning steps for each dataset to attain a higher pruning ratio, resulting in significant costs for large language models.

Other studies \cite{sanger, leopard} have also proposed token pruning methods, showing decent performance gains in bidirectional language models \cite{bert,albert}. However, they are not optimized for the memory-bounded generation phase as they require loading all \textit{KV} pairs on-chip. Thus, previous methods become suboptimal in generation as they do not adjust the pruning method for individual instances or reduce \textit{KV} transfers.

\section{Proposed Work} \label{sec:proposed_work}

\begin{figure*}[t!]
  \centering
  \includegraphics[width=\linewidth]{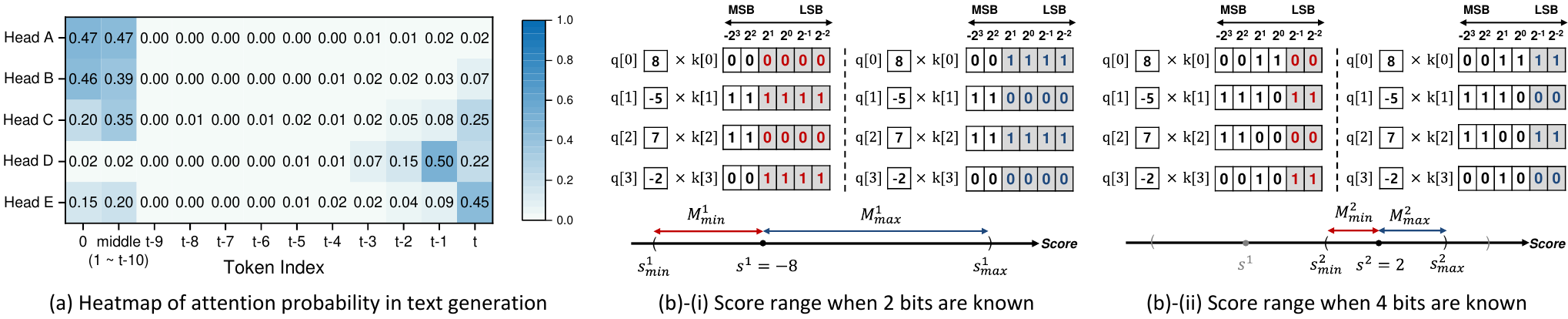}
  \vspace{-7mm}
  \caption{(a) Heatmap of attention probability across token indices in text generation, where the middle column aggregates probabilities for tokens from 1 to t-10. (b) Margins from partial score where true result exist. $s^b$ indicates partial score of chunk index $b$. $M^b_{min}$ and $M^b_{max}$ imply margins for the minimum and maximum values, respectively. }
  \label{fig:probability_heatmap_and_margin}
  \vspace*{-2mm}
\end{figure*}

To counteract the described variability, removing only the tokens with low probability is practical instead of removing them under fixed ratio. However, since the probability is determined by the difference between scores, it requires all correlation scores, which hinders the reduction of \textit{K} transfers. To identify negligible tokens and reduce \textit{K} transfers simultaneously, we propose a method that estimates the probabilities prior to completing the correlation score calculations. 

Specifically, this method estimates the probability using the segmented bits (bit chunks) of \textit{K}, and dynamically prunes tokens with the estimated value falling below a set threshold, \textit{thr} (Sec. \ref{subsec:probability_estimation}). 
If the pruning occurs before requesting the final chunk, the remaining $K$ bit chunks and $V$ vector transfer can be avoided; otherwise, the token is considered pivotal and the subsequent chunk is requested for more precise pruning decisions. In this procedure, there are two main challenges: 1) Errors by estimation lead to incorrect decisions (i.e. a pruned token has a probability larger than \textit{thr}), potentially damaging accuracy (Sec. \ref{subsec:probability_estimation}). 2) On-demand DRAM requests for bit chunks, unlike on-chip memory accesses, incur significant latency. This leads to the under-utilization of compute units (Sec. \ref{subsec:out-of-order_score_calculation}).

\subsection{Probability Estimation} \label{subsec:probability_estimation}
The nature of the exponential allows estimating probability $p_i'$ as $\frac{exp\left(s_i\right)}{\sum_{j\in subset}{exp{\left(s_j\right)}}}$ when only a \textit{subset} of tokens is known. Since the result of exponentiation is always positive, the inequality $p_i < p_i'$ holds. Therefore, if an estimated probability $p'_i$ is below a predefined $thr$, the actual probability $p_i$ will also be below, irrespective of any forthcoming scores for tokens $j \notin subset.$


To find negligible tokens as early as possible, it is advantageous to prioritize dominant tokens within the subset. To this end, we exploit the locality of attention mechanism observed in text generation. As shown in Fig. \ref{fig:probability_heatmap_and_margin}(a), recently generated tokens and the first token often carry more weights than others. Therefore, beginning the score calculation with these tokens and progressing in reverse chronological order effectively enhances the pruning ratio.

To further reduce \textit{K} transfers, we introduce a method that estimates the probability of a token using bit chunks of \textit{K}. This method extends the \textit{conservative margin} concept \cite{leopard} for 2's complement number format. The margin represents the potential amount of change. For an N-bit integer $a_{N-1}a_{N-2}\ldots a_0$, its value $w$ is:
\vspace*{-1mm}
\begin{equation}
    w = -a_{N-1} 2^{N-1} + \sum_{i=0}^{N-2} a_i 2^i
\end{equation}
In this format, all bits except the sign bit contribute a value of positive or zero. 
In a dot-production of two vectors, we can predict possible range of the result even if only a portion of bits of one operand is given.
Fig. \ref{fig:probability_heatmap_and_margin}(b) details this concept. In this example, while $Q$ retains all the bits, $K$ has a fraction of the 6 bits: 2 bits in (a) and 4 bits in (b). For elements of $Q$ that are positive, setting the unknown bits of $K$, shown in gray, to 1 (or 0 if negative) yields potential maximum score $s^b_{max}$ since it considers only increments. 
Conversely, flipping the unknown bits determines the minimum score $s^b_{min}$. Note that the margin pairs for each chunk index are determined solely by the $Q$ vector. Using this concept, in expression $\frac{exp\left(s_i\right)}{\sum_{j\in subset}{exp{\left(s_j\right)}}}$, replacing the correlation score in the numerator with $s^b_{max}$ and the score in the denominator with $s^b_{min}$ yields maximally estimated probability $p_i''$ by following relation:
\vspace*{-1mm}
\begin{equation}
    \frac{exp\left(s_i\right)}{\sum_{j\in subset}{exp{\left(s_j\right)}}} 
    \le \frac{exp(s_{i,max}^{b})}{\sum_{j\in subset}{exp(s^{b}_{j, min})}}=p_i''
\end{equation}

\begin{figure}[t!]
  \centering
  \includegraphics[width=\linewidth]{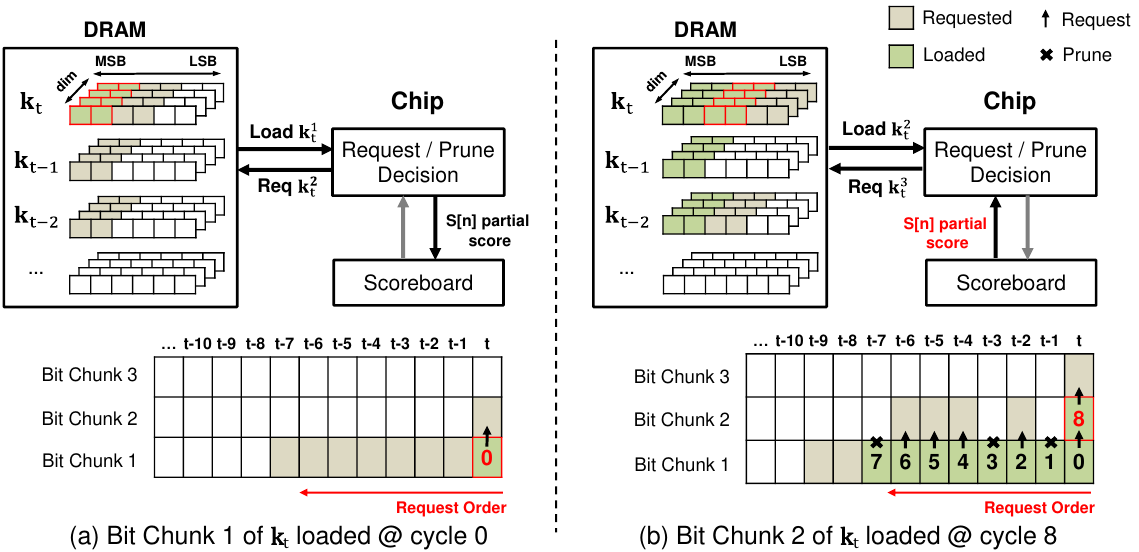}
  \vspace{-6mm}
  \caption{Out-of-Order Score Calculation}
  \label{fig:Out_of_order}
  \vspace{-5mm}
\end{figure}

Thus, if $p_i''$ falls below the $thr$, it is inferred that the $p_i$ is as well. This method ensures the safe elimination of \textit{KV} transfers of tokens with low probabilities.

In conclusion, our approach provides a conservative probability estimate. It identifies redundant tokens based on their relative importance compared to the subset, thus ensuring that all necessary tokens are retained in each instance.

\subsection{Out-of-order Score Calculation} \label{subsec:out-of-order_score_calculation}

Our estimation method permits pruning decisions that rely on chunks of \textit{K}. When a token is not pruned, the following chunk of the \textit{K} is requested to DRAM for a more precise estimation. However, waiting for each request whenever it occurs leads to under-utilization due to off-chip memory accesses latency. To mitigate this, we introduce an out-of-order computing strategy, depicted in Fig. \ref{fig:Out_of_order}, which processes as follows:

(1) When the score computation starts, only the first chunks of $K$ vectors are requested in sequence. 

(2) Once any chunk is loaded from the DRAM, the partial score of the chunk is computed, the probability is estimated, and the decision on pruning is made. 

(3) If not pruned, the next chunk of that \textit{K} is requested. Concurrently, its partial score is stored in the Scoreboard (Fig. \ref{fig:Out_of_order}(a)). Otherwise, the process of requesting the first chunk continues. 

(4) When any downstream chunk is loaded from DRAM  (Fig. \ref{fig:Out_of_order}(b)), it fetches the previous partial score from the Scoreboard, followed by an update of the partial score. Then, it repeats the process (2) and (3) with the new score. 

This out-of-order processing approach enables calculating the partial scores for different $K$s as soon as any chunk becomes available from DRAM. It facilitates the continuous score calculation through ongoing requests. For example, consider the period between the request (Fig. \ref{fig:Out_of_order}(a)) and the on-chip loading (Fig. \ref{fig:Out_of_order}(b)) of the second chunk of $\mathbf{k}_t$. During this interval, the first chunk of the other tokens is processed, which leads to either a request for a new first chunk or for the next chunk of non-pruned tokens. Therefore, this method keeps the Processing Element (PE) active and optimizes on-demand DRAM access, leading to faster execution with fewer bit accesses for pruned tokens.

This process seamlessly supports on-demand access to chunks of $K$, resulting in both speedup and reduced access. In the end, only the tokens that have not been removed by the last chunk participate in subsequent softmax and $\times V$ operations.

\vspace{-1mm}
\section{ToPick Architecture} \label{sec:topick_arch}

\begin{figure}[t]
  \centering
  \includegraphics[width=\linewidth]{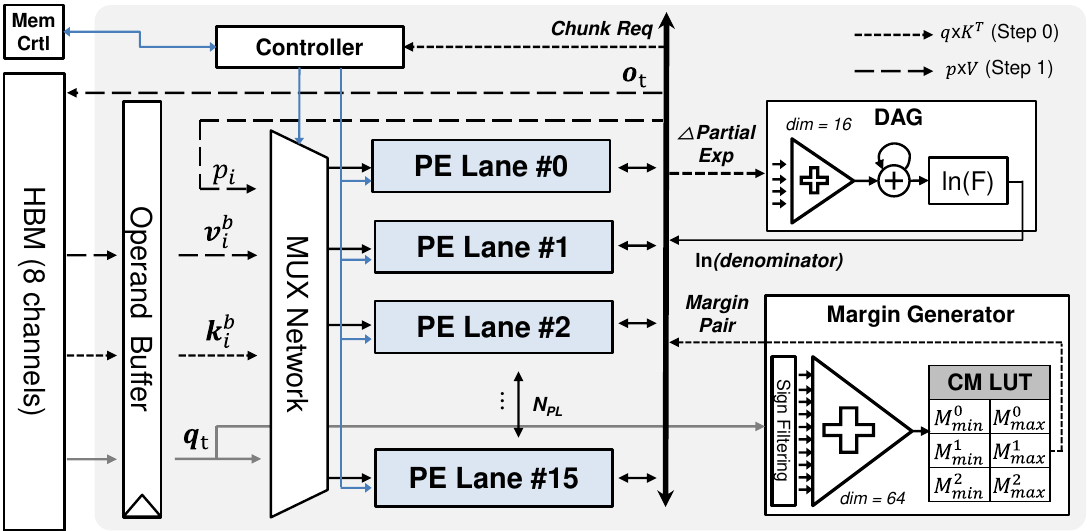}
  \vspace{-5mm}
  \caption{ToPick Overall Architecture}
  \label{fig:overall_arch}
  \vspace*{-5mm}
\end{figure}

This section presents the ToPick architecture (illustrated in Fig. \ref{fig:overall_arch}), which implements self-attention mechanism with dedicated modules: the Denominator Aggregation Module (DAG) and the Margin Generator, both supporting probability estimation and out-of-order score calculation. We design a lane-based processing element (PE Lane) for ToPick, where $K/V$ vectors from the DRAM are partitioned to each PE Lane. The operand precision for self-attention is set to 12 bits, segmented into three 4-bit chunks for each $K/V$ vector. 

ToPick performs two main operations: $\mathbf{q}_t^{T} \cdot \mathbf{k}_i$ (step 0) and $\sum p_i\mathbf{v}_{i}$ (step 1), where $i$ indicates the index of a token. The MUX network in Fig. \ref{fig:overall_arch} configures the datapath for each stage, enabling dot-product calculations for step 0 and accumulative operations for step 1. During the prompt phase, all \textit{K/V} vectors are preloaded into the on-chip buffer to be reused across queries.  Conversely, in the generation phase, the $\mathbf{q}_t$ resides in the operand buffer and each \textit{KV} chunk is streamed from DRAM, resulting in the execution of 12$\times$4 bit operations. Following are the details about hardware modules utilized for probability estimation in the generation phase:

(1) Before starting step 0, the Margin Generator produces three margin pairs ($M^b_{min}, M^b_{max}$) for each chunk index $b$ solely from the query. These margin pairs are then utilized during step 0 through a Look-Up Table (LUT) to support probability estimation.

(2)  Equipped with 64 multipliers and an adder tree, PE Lanes perform dot-products and probability estimation with out-of-order execution to minimize data transfer in step 0. Following this, in step 1, they compute attention probabilities and request $\textbf{v}_i$ for the unpruned tokens to make attention output $\textbf{o}_t$. Further details on the PE Lane design for reducing data transfer will be provided in the subsequent subsection.


(3) DAG determines a real-time denominator of $p''_i$. During each cycle, the differences between chunk indices ($\exp⁡(s_{min,i}^b ) - \exp⁡(s_{min,i}^{b-1})$) from PE Lanes are aggregated to update the denominator, ${\sum_{j\in subset}{\exp(s^{b}_{j, min})}}$. The natural logarithm of the denominator is distributed to all PE Lanes, facilitating the evaluation of $s^b_{i,max}$ - $\ln(denominator)$ $\leq$ $\ln(thr)$, which is equivalent to $p'' \leq thr$. Following step 0, the denominator represents the exponentiated sum of the unpruned scores, which is utilized for the softmax operation.


\begin{figure}[t]
  \centering
  \includegraphics[width=\linewidth]{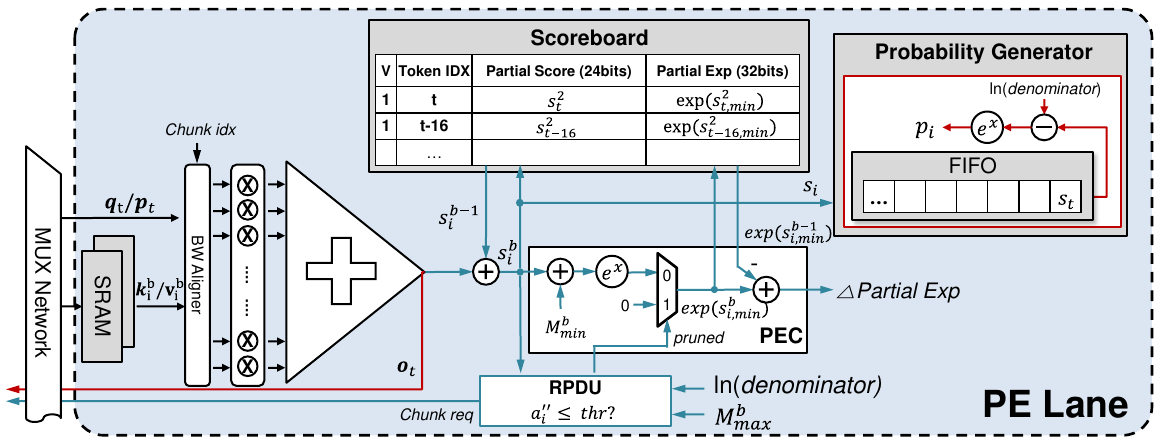}
  \vspace{-5mm}
  \caption{PE Lane Microarchitecture}
  \label{fig:microarch}
  \vspace{-5mm}
\end{figure}

\subsection{Microarchitecture of PE Lane}



Fig. \ref{fig:microarch} illustrates the PE Lane microarchitecture. It consists of four components, including the multiplier-adder tree and three modules supporting the probability estimation in out-of-order processing: (1) Scoreboard in each lane acts as temporary storage for buffering the partial results. The results include partial score $s_i^b$ and the partial exp, the exponent value of $s^b_{i,min}$ (i.e., $s^b_i + M^b_{min}$), of unpruned tokens. (2) Request/Prune Decision Unit (RPDU) makes a prune decision and decides which chunk to request. (3) Partial Exp Calculator (PEC) makes the partial exp to aggregate the denominator. If the chunk is downstream one, calculates the difference of partial exp between chunks.

These modules jointly operates to make prune decision and creating the partial exp for aggregating the denominator in step 0. At first, the multipliers and adder-tree computes the dot-product result $ps_i^b$ from a 12-bit vector $\mathbf{q}_t$ and a 4-bit chunk vector $\mathbf{k}^b_i$. At the same time, the Scoreboard is accessed with token index $i$. If previous chunk exist, the previous score $s^{b-1}_{i}$ is fetched and updated, generating new partial score $s^{b}_{i}$. 

After that, prune decision is determined in RPDU. The RPDU gets the upper margin $M^b_{max}$, $\ln(denominator)$, and the partial score $s^{b}_{i}$. Then the unit determines  $p_i'' \leq thr$  holds. If true, the unit requests new first chunk to DRAM. Otherwise, it requests the subsequent chunk vector for the $\mathbf{k}_i$ and store the partial results in Scoreboard.



PEC makes the partial exp that is $\exp(s^b_{i,min})$. It generates a difference of exponent values between the chunk index to deduct the previous value $\exp(s^{b-1}_{i,min})$ from the denominator. All the difference values collected from PE Lanes are aggregated in the DAG.

\begin{figure*}[ht!]
  \centering
  \includegraphics[width=\linewidth]{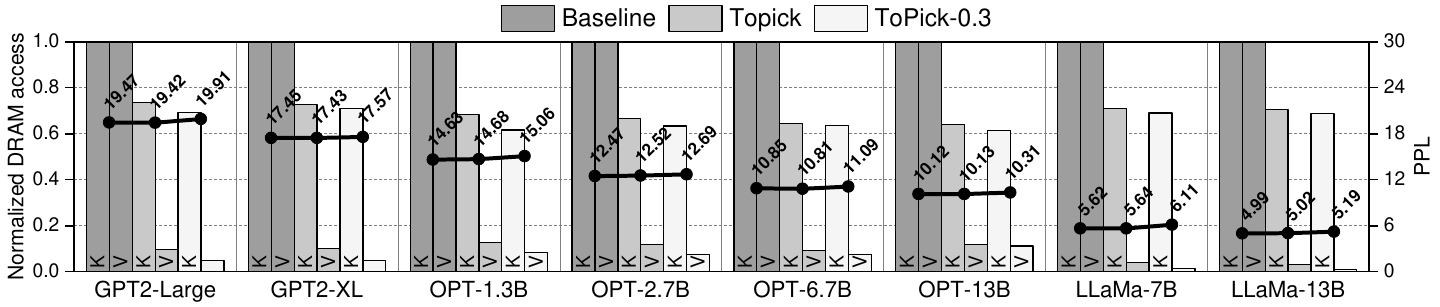}
  \vspace{-6mm}
  \caption{Required off-chip memory access in generation phase (bars) and algorithm performance (lines) across varying models.}
  \Description{Impact of proposed on performance across varying target probability (\textit{thr}).}
    \label{fig:KV_reduction}
  \vspace*{-4mm}
\end{figure*}

The Probability Generator computes attention probabilities for unpruned tokens. After step 0, the FIFO buffers indices and scores of these tokens. It calculates the probability $\exp(s_i$ – $\ln(denominator))$ that is equivalent to $p_i$. Simultaneously, it requests the corresponding $\mathbf{v}_i$ vector from DRAM. The calculated probabilities, $p_i$, are then forwarded to the PE lane, where the multiplier-adder tree performs the weighted sum $\sum p_i\mathbf{v}_{i}$, forming the attention output $\mathbf{o}_t$.


\section{Experiments} \label{sec:experiments}

\subsection{Experimental Setup} \label{subsec:setup}
\subsubsection{Algorithm Evaluation Setup}
We evaluate our proposed meth-od on various language models tailored for text generation: GPT2-Large/XL \cite{gpt2}, OPT-1.3B/2.7B/6.7B/13B \cite{opt}, and LLaMa-2-7B/13B \cite{llama2}. To analyze the impact of our methods on model performance, we measure perplexity (PPL) on the Wikitext-2-raw dataset \cite{wikitext2}, where lower perplexity values indicate better performance. This assessment utilizes pre-trained models available on Huggingface \cite{huggingface}. 

\subsubsection{Hardware Evaluation Details}
Table 1 shows the hardware configuration of the ToPick architecture. We set the number of PE Lanes to 16 to fully utilize DRAM (HBM2) bandwidth in the generation phase, where each Lane processes a chunk (4 bits) of a $K$ vector per cycle. We implement ToPick in RTL and synthesize it using Synopsys Design Compiler under Samsung 65nm LP standard cell library to evaluate the area and power consumption of ToPick at a target frequency of 500MHz (Table 2). We also use CACTI \cite{cacti} to estimate the energy and area of on-chip buffers and scoreboard. To get the number of cycle and energy of off-chip accesses, we use DRAMsim3 \cite{dramsim3} simulator with trace files generated in RTL simulation.

\begin{table}[h]
\newcolumntype{Y}{>{\centering\arraybackslash}m{0.25\linewidth}}
\newcolumntype{Z}{>{\raggedright\arraybackslash}m{0.7\linewidth}}

\vspace*{-3mm}
\caption{Hardware Configurations of ToPick}
\vspace*{-3mm}
\centering
\small
\begin{tabularx}{\columnwidth}{Y|Z}
\Xhline{3\arrayrulewidth}
Main Memory & HBM2; 8 channels $\times$ 128-bit at 2GHz;
\newline each channel provides 32GB/s bandwidth.\\ 
\hline
On-chip Buffer & 192KB SRAM for each Key, Value buffer;
\newline 512B Operand buffer.\\ 
\hline
PE Lane  & 64-dim $\times$ 12-12 bit multipliers and adder tree; 
\newline 32 entry $\times$ 67 bit Scoreboard;
\newline 2 $\times$ 32 bit fixed-point EXP unit. \\ 

\Xhline{3\arrayrulewidth}
\end{tabularx}

\vspace*{-6mm}
\end{table}

\subsubsection{Design Configurations}
To assess the efficacy of our proposed method in the generation phase, we compared our design with a baseline accelerator that lacks five hardware modules: Margin Generator, DAG, PEC, Scoreboard, and RPDU, which are integral to proposed optimizations. We evaluate two configurations with the baseline, ToPick and ToPick-0.3. The ToPick configuration includes modules supporting our methods, showing a minimal performance decrease of at most +0.05 PPL. In contrast, ToPick-0.3 is a configuration designed to balance hardware benefits with a slight performance decrease, allowing for increase of +0.3 PPL on average in Wikitext-2. For all hardware evaluations, we use context length of 1024 for GPT2 models and 2048 for OPT, LLaMa-2 models.

\subsection{Result}




\subsubsection{Memory access reduction} 
\renewcommand{\thefootnote}{\arabic{footnote}}
\begin{figure}[t!]
  \centering
  \includegraphics[width=\linewidth]{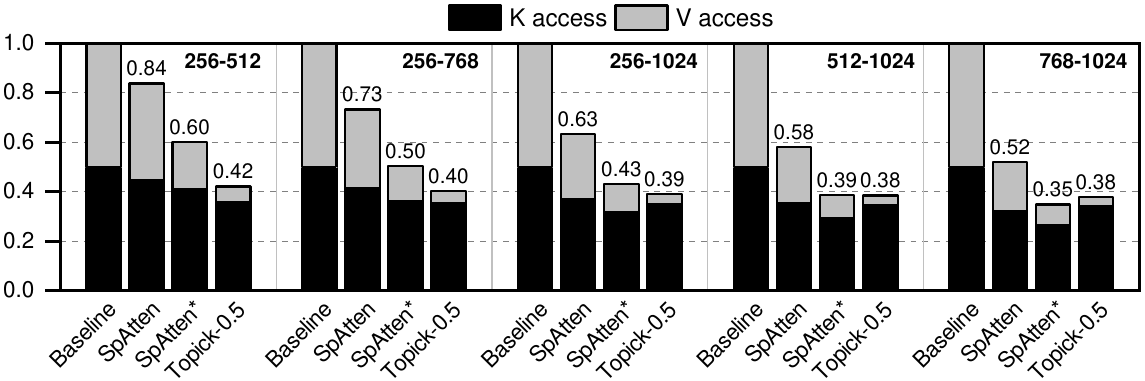}
  \vspace{-6mm}
  \caption[h]{Normalized memory access comparison. \hyperlink{footnote1}{\protect\footnotemark[1]} \hyperlink{footnote1}{\protect\footnotemark[2]}}
  \label{fig:comparison}
  \vspace*{-6mm}
\end{figure}

\hypertarget{footnote1}{}
\footnotetext[1]{ SpAtten* performs additional fine-tuning on Wikitext-2 dataset.\label{footnote1}}
\footnotetext[2]{ The notation "a-b" for each cell indicates the prompt length and the ending length in text generation.}

\begin{figure*}[t]
  \centering
  \includegraphics[width=\linewidth]{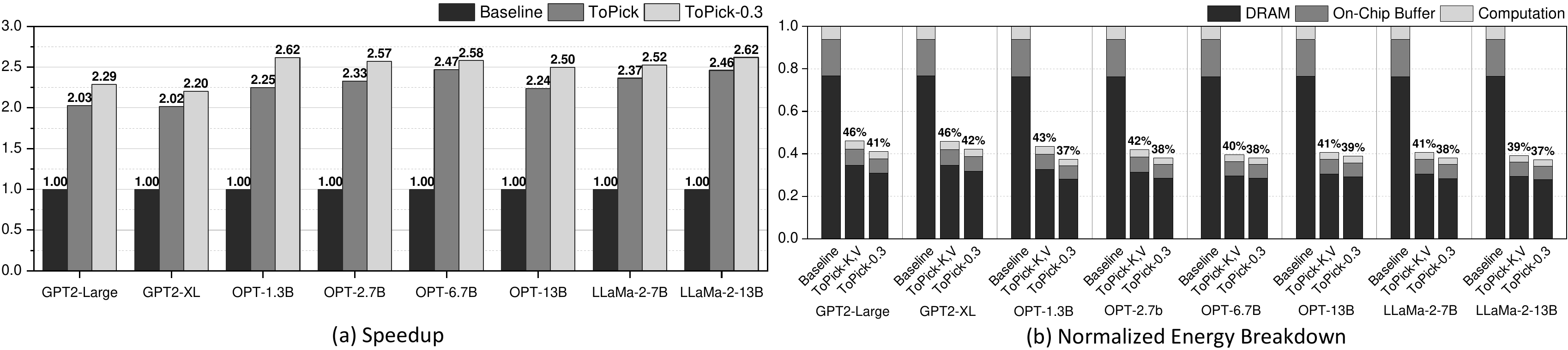}
  \vspace*{-2mm}
  \caption{(a) Normalized energy breakdown and (b) Speedup of ToPick configurations in generation phase.}
  \label{fig:hardware_eval}
  \vspace*{-2mm}
\end{figure*}

Fig.\ref{fig:KV_reduction} illustrates the impact of our method on the reduction of off-chip access for \textit{KV caching} in the generation phase, showing a normalized comparison to the baseline. Our probability estimation scheme selectively prunes only the unimportant tokens in each instance, achieving a 12.1$\times$ and 22.2$\times$ reduction in \textit{V} access in ToPick and ToPick-0.3 configurations, respectively. Moreover, the out-of-order score calculation method effectively eliminates the need to load remaining chunk of \textit{K} for pruned tokens, resulting in a reduction of \textit{K} accesses by 1.45$\times$ and 1.51$\times$ on average. As a result, our method achieves a total of 2.57$\times$ and 2.79$\times$ off-chip memory access reduction in each configuration.

Fig. \ref{fig:comparison} compares of ToPick-0.5 and Spatten \cite{spatten} across various context lengths using the GPT2-Medium \cite{gpt2} model. For a fair comparison, we set the precision of \textit{Q}, \textit{K}, \textit{V} to 12 bits and allow a +0.5 PPL in the Wikitext-2 dataset for both configurations. 
As shown in the figure, our scheme generally shows a better reduction ratio than SpAtten, except in the case of longer prompt setting where the cascaded token/head pruning method significantly reduces \textit{K} access. However, when evaluating both designs without fine-tuning, our scheme shows a 1.64$\times$ higher reduction in memory accesses compared to SpAtten.

\subsubsection{Speed up \& Energy Efficiency in Generation Phase}
Fig. \ref{fig:hardware_eval} presents the (a) speedup and (b) normalized energy breakdown of our ToPick accelerator during the generation phase for various models. As outlined in Sec. \ref{subsubsec:KV_caching}, the workload of the generation phase is memory-bounded; in the baseline design, latency and energy consumption are primarily due to off-chip memory accesses. 

The introduction of probability estimation substantially reduces the access to \textit{V}, yielding a speedup of 1.73$\times$ and energy savings of 1.78$\times$ compared to baseline. These results are due to the precise prediction of low probability tokens. Integrating out-of-order score calculation, which is ToPick, leads to further benefits---a 1.32$\times$ increase in speed and 1.35$\times$ saving in energy consumption on average. The out-of-order technique hides the latency of on-demand DRAM access via independent score computations, thus accelerating the process.  Moreover, by accepting a minor algorithmic degradation, ToPick-0.3 achieves a speedup of 2.48$\times$ and energy efficiency of 2.63$\times$. These results underscore the effectiveness of our proposed design in optimizing self-attention execution for text generation.


\begin{table}[h!]
\newcolumntype{Y}{>{\centering\arraybackslash}m{0.24\linewidth}}
\newcolumntype{Z}{>{\centering\arraybackslash}m{0.11\linewidth}}

\vspace*{-3mm}
\caption{Area and Power breakdown of ToPick at 500MHz.}
\label{tab:Area_power}
\vspace*{-3mm}
\centering
\begin{tabularx}{\columnwidth}{Z|Y|Y|Y}
\Xhline{3\arrayrulewidth}
\multicolumn{2}{c|}{\textbf{Hardware Module}} &  \textbf{Area (mm${^2}$)} & \textbf{Power (mW)}  \\ 
\hline \hline
\multicolumn{2}{c|}{\small{PE Lane $\times$ 16}} &  2.518   & 426.76 \\ 
\hline
\multirow{5}{*}{\small{PE Lane}} & \small{Multipliers \& Adder-Tree 12b} &  0.095   & 17.94  \\
\cline{2-4}
& \small{Prob Gen} &  0.032   & 2.22 \\
\cline{2-4}
& \small{PEC} &  0.004   & 0.73 \\ 
\cline{2-4}
& \small{Scoreboard} &  0.024   & 4.69 \\
\cline{2-4}
& \small{RPDU} &  0.001   & 0.17 \\
\hline
\multicolumn{2}{c|}{\small{Mux Network}}   & 0.076   & 3.13  \\ 
\hline
\multicolumn{2}{c|}{\small{Margin Generator} }  & 0.014   & 3.78  \\
\hline
\multicolumn{2}{c|}{\small{DAG}}   & 0.010   & 2.49  \\
\hline
\multicolumn{2}{c|}{\small{On-chip buffer}}   & 5.968   & 1053.32  \\
\hline
\hline
\multicolumn{2}{c|}{\small{Total} }  & 8.593   & 1492.78  \\
\Xhline{3\arrayrulewidth}
\end{tabularx}

\vspace*{-3mm}
\end{table}

\subsubsection{Area \& Power Analysis}

Table \ref{tab:Area_power} presents the area and power analysis for the ToPick accelerator. The additional modules, Margin Generator, DAG, and PEC, are designed to minimize \textit{V} access on top of the baseline configuration, resulting in an area overhead of 1.0\% and a power overhead of 1.3\%. The ToPick further incorporates modules, Scoreboard and RPDU, aimed at reducing \textit{K} access. This results in an additional area and power overhead of 4.9\% and 5.6\%, respectively, over the baseline. Nevertheless, the increase in power consumption is alleviated by significant power savings from the reduction of off-chip memory accesses. Therefore, the ToPick architecture demonstrates a strategic compromise, exchanging a moderate rise in power for a substantial improvement in energy efficiency.

\section{Conclusion}
In this work, we address the \textit{KV caching} transfer overhead of self-attention in text generation. 
Firstly, we propose a probability estimation that identifies unimportant tokens using partial bits of \textit{K}. 
This method prunes negligible words aligned with each instance, achieving a substantial pruning ratio for \textit{V} without fine-tuning, while also creating an opportunity to reduce \textit{K} access.
Secondly, we design an architecture that supports seamless on-demand off-chip DRAM access.
Through out-of-order execution, our design avoids under-utilization by ongoing DRAM request. 
In conclusion, our proposed method reduces 2.57$\times$ off-chip DRAM access for self-attention in text-generation, realizing a speedup of 2.28$\times$ and 2.41$\times$ energy efficiency.


\bibliographystyle{ACM-Reference-Format}
\bibliography{sample-base.bib}
\end{document}